\documentclass[a4paper,11pt]{article}
\usepackage{pos}
\usepackage{slashed}

\newcommand*{\no}{\noindent}
\newcommand*{\bea}{\begin{eqnarray}}
\newcommand*{\eea}{\end{eqnarray}}
\newcommand*{\be}{\begin{equation}}
\newcommand*{\ee}{\end{equation}}

\newcommand*{\pref}[1]{(\ref{#1})}

\newcommand*{\str}{\mathrm{str}}

\newcommand*{\prefr}[2]{(\ref{#1}-\ref{#2})} 
\newcommand*{\nn}{\nonumber}

\newcommand*{\tl}{\mathrm{tl}}

\newcommand{\bma}{\begin{pmatrix}}
\newcommand{\ema}{\end{pmatrix}}

\newcommand*{\la}{\left\langle}
\newcommand*{\ra}{\right\rangle}

\title{Subleading Higgs effects at lepton colliders}

\author*[a]{Axel Maas}
\author[b]{Duifje M.\ van Egmond}
\author[a,c]{Simon Pl\"atzer}

\affiliation[a]{Institute of Physics, NAWI Graz,\\
  Universit\"atsplatz 5, 8010 Graz, Austria}
\affiliation[b]{Instituto de Física, Universidade Federal Fluminense,\\
Campus da Praia Vermelha, Av. Litorânea s/n,\\
24210-346, Niterói, RJ, Brasil}
\affiliation[c]{Particle Physics, Faculty of Physics,\\
University of Vienna, Boltzmanngasse 5,\\
A-1090 Wien, Austria}

\emailAdd{axel.maas@uni-graz.at}
\emailAdd{duivemaria@ictp-saifr.org}
\emailAdd{simon.plaetzer@uni-graz.at}

\abstract{Subtle field-theoretical effects suggest the presence of additional Higgs contributions in standard model processes. This has been supported by electroweak lattice calculation, e.\ g.\ for vector boson scattering. These effects can be included in perturbation theory by a suitable augmentation. We use such augmented perturbation theory to determine the impact at next-to-leading order at lepton colliders, from LEP to future machines such as FCC, in collisions with fermion-antifermion final states. After providing the formal background, we outline the calculational procedure, showing that in the fully exclusive process ${e^\text{\textendash}}e^+\to f\bar{f}$ deviations only occur in fixed order at electroweak NNLO, but become relevant at the TeV scale already in resummed tree-level calculations. We discuss further processes where deviations are expected already at fixed-order NLO.}

\FullConference{42nd International Conference on High Energy Physics (ICHEP2024)\\
18-24 July 2024\\
Prague, Czech Republic\\}

\begin{document}
\maketitle

\section{Introduction: The need to augment perturbation theory}

Phenomenology using perturbation theory has been tremendously successful for electroweak physics \cite{pdg}. However, ever since the seminal work of Fr\"ohlich, Morchio, and Strocchi (FMS) almost half a century ago \cite{Frohlich:1980gj,Maas:2017wzi}, is was known that this is due to a non-trivial interplay of the group-theoretical structure of the standard model and the Brout-Englert-Higgs (BEH) effect. In actuality, gauge symmetry requires also a description of weakly charged states, especially in the initial state and final state in experiments, in a composite way. But the structural and kinematic features of the standard model suppress deviations between such a full description and a perturbative one substantially. While there are necessarily deviations eventually, it is yet quantitatively unclear in which processes they will surface to which extent \cite{Fernbach:2020tpa,Jenny:2022atm,Maas:2022gdb}.

Such deviations could be, and have been \cite{Maas:2017wzi,Jenny:2022atm}, described in full non-perturbative investigations. It is possible to circumvent this formidable challenge, however, with the aid of the Brout-Englert-Higgs effect. Expanding composite states in terms of the Higgs vacuum expectation value, the so-called FMS mechanism \cite{Frohlich:1980gj,Maas:2017wzi} allows to augment usual perturbation theory to capture the additional contributions to a large extent \cite{Maas:2017wzi,Dudal:2020uwb,Maas:2020kda}.

In the following, augmented perturbation theory for scattering processes will be introduced briefly. It will then be shown that in the exclusive inelastic process ${e^\text{\textendash}}e^{+}\to f\bar{f}$, with $f\neq e$ arbitrary fermions, first corrections from augmentation start only from NNLO electroweak, and the NLO electroweak augmentation contributions vanish exactly. This explains why the effect has not been seen in this standard candle process at  past lepton colliders. This discussion will be wrapped up with some remarks on how resummation can change this, especially at future high-energy lepton colliders, and in which processes contributions may already surface at NLO electroweak.

\section{Augmenting perturbation theory}

The basic tenet of the FMS construction \cite{Frohlich:1980gj} is that electroweak gauge symmetry breaking is not literally there. This follows from field-theoretical arguments, e.\ g.\ Elitzur's theorem and its generalizations \cite{Frohlich:1980gj,Maas:2017wzi}. Thus, physical states need to be manifestly weakly gauge-invariant. The role of the BEH effect in this context is subtle, but basically guarantees that an augmentation of perturbation theory is sufficient, rather than requiring a full non-perturbative treatment, as in QCD. A full discussion of the theoretical background and repercussions, as well as an overview of the extensive body of varied support of this scenario, can be found in the reviews \cite{Maas:2017wzi,Maas:2023emb}.

As a consequence, the external states in a scattering process need to be described in terms of gauge-invariant composite operators, rather than elementary ones. A physical left-handed lepton is specified by its generation $g$, its physical flavor $a$, and its spin state $\lambda$. Such a state is described by the composite operator $\Psi^{ag}_\lambda=X_i^{a \dagger}\psi_{L\lambda}^{gi}$, where $i$ is the contracted weak gauge index. Herein $\psi_L$ is the left-handed elementary lepton fields and $X$ a matrix-valued representation of the Higgs field \cite{Maas:2017wzi}. Note that physical flavor $a$ refers to the global (custodial) symmetry of the Higgs field, which is explicitly broken by the Yukawa interactions. Hypercharge is left implicit, as it will not play a role here. A detailed discussion of it can be found in \cite{Maas:2017wzi}. The Higgs field $X$ is the one before splitting off the vacuum expectation value, to maintain exact weak gauge invariance. Thus, splitting in the composite operator $X$ as $X=v+\eta$, with $v$ the Higgs vacuum expectation value, yields that the operator up to Higgs fluctuation effects behave like the elementary fermion operator of perturbation theory \cite{Frohlich:1980gj,Maas:2017wzi}. The smallness of the Higgs fluctuations due to the BEH effect is the origin of the suppression of the remainder term. This is the FMS mechanism.

Thus, any scattering process involving left-handed fermions has, in principle, to be described in terms of the scattering of composite states, very much like the description of hadron scattering. The corresponding composite LSZ formalism is well understood \cite{Meissner:2022cbi,MPS:unpublished}. Making it tractable using perturbative means, rather than full non-perturbative ones \cite{Jenny:2022atm}, is where FMS-augmented perturbation theory comes into play \cite{MPS:unpublished}. This will be elucidated now.

Consider as the external state a left-handed electron. Applying the FMS mechanism to the original connected matrix element yields
\bea
&&{\cal M}^{\Psi^{ag}_\lambda...}(P...)=\int \frac{d^4q}{(2\pi)^4}\la X^{a\dagger}_i\psi_{L\lambda}^{ig}(P-q)...\ra_c\nn\\
&=&v^{ai}\la \psi^{ig}_{L\lambda}(P)...\ra_c+\int \frac{d^4q}{(2\pi)^4}\la \eta^{ai}(q)\psi^{gi}_{L\lambda}(P-q)...\ra_c\nn,
\eea
\no where the integration over relative momenta stems from having a composite operator. The first term is the one of usual perturbation theory, up to a prefactor. This prefactor projects the full elementary fermion field on the chosen direction of the vacuum expectation value. Choosing it, e.\ g., in the conventional three-direction, this reduces to the very same matrix element as in ordinary perturbation theory. The second term involves a composite operator, now with the fluctuation Higgs field. In case of a right-handed electron this term would be absent.

Applying the LSZ construction to the first term works as usual. Applying it to the second term can be done by means of Bethe-Salpeter amplitudes \cite{Meissner:2022cbi,MPS:unpublished}. The matrix element then becomes
\be
{\cal M}^{\Psi^{ag}_\lambda...}_\text{LSZ}(P...)=Z^\frac{1}{2}v^{ai}\la \psi^{ig}_{L\lambda}(P)...\ra_\text{truncated}+\int \frac{d^4q}{(2\pi)^{4}}\bar{\chi}^{abgh}_{ij\lambda\alpha}(P,q)\times\la \eta^{bj}(q)\psi^{hi}_{L\alpha}(P-q)...\ra_\text{truncated}\label{truncated}
\ee
\no where $Z$ is a suitable wave-function renormalization and $\chi$ is the Bethe-Salpeter amplitude. The truncation of the matrix element occurs on the level of the four-point function $G_4$ connecting to the composite operator. However, perturbatively symbolically
\be
G_4^{-1}=(1+\alpha)^{-1}\approx 1-\alpha\nn
\ee
\no and thus symbolically
\be
G_4^{-1}G_4\approx(1-\alpha)(1+\alpha)=1-\alpha^2\approx 1\nn.
\ee
\no Hence, truncating the matrix element involving the composite operator is done by removing all diagrams from the full matrix elements, which are 2PI with respect to the composite operator. These describe interactions of the two constituents before they interact with other particles. The corresponding contributions are all subsumed in the Bethe-Salpeter amplitude. The Bethe-Salpeter amplitude is given by the three-point vertex connecting the composite state to the constituents as \cite{Alkofer:2000wg}
\be
\chi^{abgh}_{ik\lambda\alpha}(P,q)=\Gamma^{\Psi\psi \eta}_{gkacmn\lambda\beta}(P,q)D^\eta_{cbmi}(q)D^\psi_{klnj\beta\alpha}(P-q)\nn
\ee
\no with the propagators of the constituent particles $D$.

The truncated vertex $\Gamma$ is defined in terms of the truncated four-point scattering amplitude $\Gamma_4$
\bea
&&\Gamma_4^{aijg\lambda\to bklh\alpha}(P,P-q,P,P-k)\nn\\
&=&\Gamma^{aijg\lambda\to c m\beta}(P,P-q)\frac{\delta_{cd}\delta_{mn}\left(\slashed{P}-M\right)}{P^2-M^2+i\epsilon}\Gamma^{dn\beta\to bklh\alpha}(P,P-k)+\text{regular terms at $P^2=M^2$}\nn,
\eea
\no where $M$ is the mass of the desired external state, in the present case the electron. The relevant three-point vertex is
\be
\Gamma^{aijg\lambda\to c m\beta}=\la \bar{\Psi}^{cm}_\beta(P)\psi_{L\lambda}^{gj}(P-q)X_{ai}(q)\ra_\text{connected, truncated}\nn.
\ee
\no Applying the FMS mechanism to this vertex once more, this yields for the additional contribution
\bea
&&\int \frac{d^4q}{(2\pi)^4}v_{ak}^\dagger\Gamma_{\psi_L\to\psi_L\eta}^{gk\lambda\to ih\alpha bj}(P,P-q,q) D^{ih mk}_{\alpha\beta}(P-q)D^{bcjn}(q)\la \eta^{cn}(q)\psi^{km}_\beta(P-q)...\ra \label{i2}\\
&+&\int \frac{d^4kd^4q}{(2\pi)^8}\Gamma_{\psi_L\eta\to\psi_L\eta}^{gk\lambda ak\to ih\alpha bj}(P-k,k,P-q,q) D^{im}(P-q)D^{bcjn}(q)\la \eta^{cn}(q)\psi_i(P-q)...\ra.\nn\\
\label{i3}
\eea
\no Herein the $\Gamma$ are perturbative vertex functions. The first term \pref{i2} contains the situation where the external leg is obtained from a fusion of the Higgs and elementary lepton into the FMS-dominant constituent, while the term \pref{i3} yields the interacting part propagating. As such, \pref{i2} therefore contains similar terms than the perturbative ones, and thus will yield a reweighting of those in the complete expression. The contributions in \pref{i3} are genuinely new. Still, the whole expression is now a closed expression, which can be treated in fixed-order perturbation theory.

\section{Colliding leptons}

\begin{figure}
 \includegraphics[width=\textwidth]{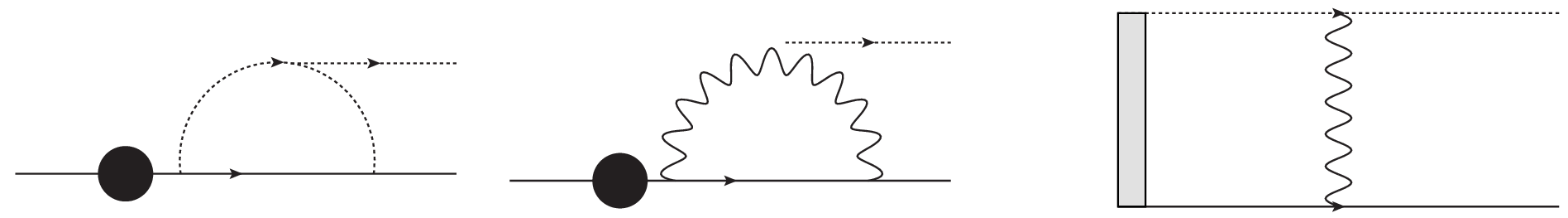}
 \caption{The leading-order diagrams contributing to $\Gamma_{\psi_L\to\psi_L\eta}$ (left two diagrams) and $\Gamma_{\psi_L\eta\to\psi_L\eta}$ (right diagram) to \prefr{i2}{i3} in the standard model. Full lines are left-handed fermions, dotted lines are Higgs fluctuations, and wavy lines are $Z$ bosons. The black dot indicates the external leg in \pref{i2} and the shaded box the additional integration in \pref{i3}.}
 \label{bse}
\end{figure}

The remainder of the calculation is a straightforward application of ordinary perturbation theory. However, because the vertices in \prefr{i2}{i3} connect left-handed fermions legs with Higgs legs, there is no tree-level contribution in the standard model. The lowest order is thus one-loop, as shown in figure \ref{bse}. Since in both cases one (in \pref{i2}) and two (in \pref{i3}) integrations are required, this makes the expression effectively two-loop and three-loop, respectively. Even though, in case of \pref{i3}, the second integration is not coming with an additional suppression by a coupling constant. Nonetheless, even at zero Yukawa coupling, and thus in the massless case, both vertices are non-vanishing.

As a consequence, in the $e^+e^-\to\bar{f}{f}$ case, the first correction appears at NNLO electroweak, and NLO electroweak is identical to the non-augmented perturbative result. This is not surprising, as both constituents of the composite operators are required to interact, which is effectively equivalent to hadronic double-parton scattering. With only two final state particles, one loop suppression is inevitable, and the second comes from the chiral structure of the standard model. Given LEP(2)'s precision, such a suppressed effect could not be detected. The question is thus, if there are processes, where the suppression can be one-loop only. There are three possibilities.

\begin{figure}
 \includegraphics[width=\textwidth]{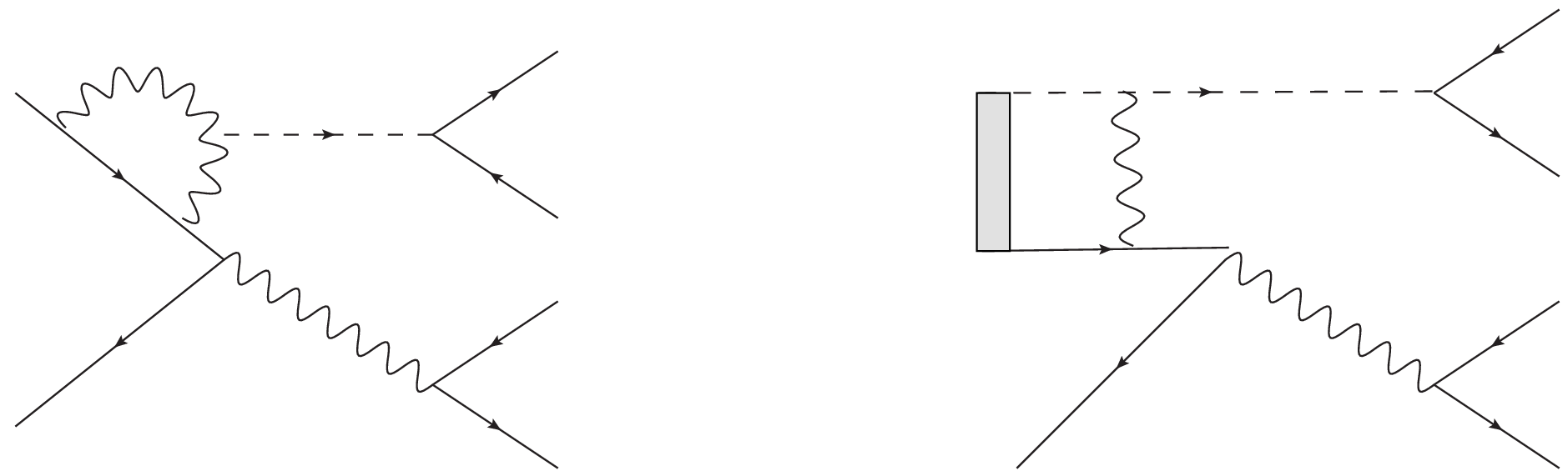}
 \caption{Leading contributions, due to \prefr{i2}{i3}, in the process $e^+e^-\to{\bar f}f{\bar F}{F}$, at zero Yukawa coupling.}
 \label{f4}
\end{figure}

One is to increase the number of external particles, e.\ g.\ in a $2\to 4$ processes. Then the second particle of the composite operators does not need to be reabsorbed, lowering the number of loops by one. An example of such a possibility is given in figure \ref{f4}, which shows a $e^+e^-\to{\bar f}f{\bar F}{F}$ process. It is visible how \pref{i2} is a process also encountered in non-augmented perturbation theory. Basically, what happens is that the composite structure allows for more possible origins of such a graph, which modifies the weight factor of the diagram in the matrix element. The second one, generate from \pref{i3}, is genuinely new. A full calculation of such a process is still needed.

A variation on this is the second option. By going to a (semi-)inclusive process, some constituents are not interacting, but need to be only summarily summed over. This would yield a PDF-like description, though with calculable PDFs in principle \cite{Egger:2017tkd}. For a predictive results, such a calculation of PDFs still needs to be done.

Resummation effects can alter the result. Now all processes need to be summed over all possible weak charges, restoring the Bloch-Nordsieck (and Kinoshita-Lee-Naunberg) theorem \cite{Maas:2022gdb}. As a consequence, double Sudakov logarithms $\sim\ln^2 s/m_W^2$ are canceled. In the processes $e^+e^-\to\bar{f}{f}$, this yields a correction of order the strong interactions contributions at the TeV scale \cite{Ciafaloni:2000rp}. Thus, again undetectable at LEP(2) energies, this would be a striking effect at future lepton colliders.

\section{Summary}

The FMS framework provides the most powerful predictive framework for understanding theories with a BEH effect. In particular, it predicts and explains the success of perturbation theory in the standard model, as well as is able to do so in cases where perturbation theory fails like (simplified) grand-unified theories, see \cite{Maas:2017wzi,Maas:2023emb} for detailed reviews. As such, it requires to go beyond perturbation theory to fully describe all features of such theory. This also applies to the standard model, where deviations in some truncated subsector have already been visible in lattice simulations \cite{Maas:2023emb,Jenny:2022atm}. The simplest, and at the same time most effective, such extension is augmented perturbation theory \cite{Maas:2017wzi,Maas:2020kda,Dudal:2020uwb,MPS:unpublished}.

Here, an outline of the steps needed to described cross sections in augmented perturbation by virtue of the LSZ construction has been given \cite{MPS:unpublished}. Applying these to a sample processes, $e^+e^-\to\bar{f}{f}$, it is seen that the effect is twice loop-suppressed. On the one hand, this clearly shows why it was not detected so far. On the other hand, this will make an experimental detection in this process also very demanding in the future. It is seen that this is an unfortunate consequence of the special structure of the standard model, making the discovery of the FMS mechanism very demanding. However, other processes have been discussed, where the suppression in the standard model may be only one loop electroweak, making a detection at a future lepton collider feasible. This will need more detailed studies to quantify.

The FMS mechanism is a unambiguous feature of the standard model. This implies that either the modifications to experimental observables this demands will be discovered eventually, or new physics beyond the standard model has to explain their absence. Given that it is currently the framework with most explanatory power for theories with a BEH effect, this appears a more than worthwhile endeavor. Not to mention its impact for the search for new physics \cite{Maas:2023emb}.

\bibliographystyle{bibstyle}
\bibliography{bib}

\end{document}